\documentclass[12pt]{article}
\usepackage{mathtools,amssymb,mathrsfs,ytableau,microtype,tikz,slashed}
\usepackage{physics}
\usepackage{color}
\usepackage[linktocpage]{hyperref}
\hypersetup{colorlinks=true,citecolor=red,linkcolor=black, urlcolor=blue, breaklinks=true}
\usepackage{cite}
\usepackage{moresize}
\usepackage{url}
\ytableausetup{centertableaux,boxsize=.5em}
\usepackage[numbers,sort]{natbib}

\makeatletter\@addtoreset{equation}{section}\makeatother

\renewcommand{\title}[1]{\vbox{\center\LARGE{#1}}\vspace{5mm}}
\renewcommand{\author}[1]{\vbox{\center\large#1}\vspace{5mm}}
\newcommand{\address}[1]{\vbox{\center\em#1}}

\begin{document}

\begin{titlepage}

\begin{center}

\vskip 0.5cm

\title{
 Generalized Symmetries Phase Transitions \\ with Local Quantum Fields
}

 \author{
 Po-Shen Hsin$^{1}$
 }

\address{${}^1$ Department of Mathematics, King’s College London, Strand, London WC2R 2LS, UK.}

\end{center}

\abstract{
Symmetries are important guiding principle for phase transitions.
We systematically construct field theory models with local quantum fields that exhibit the following phase transitions:
 (1) different symmetry protected topological (SPT) phases with generalized symmetries; (2) different symmetry enriched topological (SET) phases with generalized symmetries differ by symmetry fractionalizations; (3) spontaneously broken generalized symmetries, where the unbroken phases can have nontrivial SPT or SET. 
 The models are ordinary gauge theories with bosons or fermions in 3+1d and 2+1d. 
We focus on one-form symmetries and symmetries generated by condensation defects, which do not act on local operators. The phase transitions are protected from local operator perturbations which do not change the asymptotic phases.
In particular, we show that continuous gauge theories in 3+1d can have different phases distinguished by fractionalizations of unbroken one-form symmetries.
}

\vfill


\vfill

\end{titlepage}

\eject

\tableofcontents

\unitlength = .8mm

\setcounter{tocdepth}{3}

\bigskip

\section{Introduction}

Symmetries are important guiding principle for phases of matter: if two systems have different symmetry properties, they must be separated by phase transitions unless the symmetries are violated explicitly. This includes the Landau paradigm for spontaneous symmetry breaking, as well as topological phase transitions:
\begin{itemize}

    \item Symmetry protected topological (SPT) phase transitions. Across the transition, the systems belong to distinct SPT phases. This includes the integer quantum Hall transitions, as well as the trivial insulator/topological insulator transitions.\footnote{
    In this work, we will focus on SPT instead of gapless SPTs for simplicity. We will discuss gapless SPT \cite{PhysRevX.6.011034,Hsin:2019fhf,Thorngren:2020wet,Dumitrescu:2023hbe} transitions in a separate work.
    }

    \item More generally, there are symmetry enriched topological (SET) phase transitions. Across the transition, the system belong to distinct SETs: they can be different topological orders, or the same topological order with different symmetry fractionalizations \cite{Barkeshli:2014cna,Teo_2015,Tarantino_2016,Benini:2018reh,Hsin:2019fhf,Bhardwaj:2023wzd,Bartsch:2023pzl}. This includes fractional quantum Hall transitions. 

    \item Variants of symmetry breaking transitions between symmetry breaking phase and unbroken phase that exhibits nontrivial SPT or SET.
    
\end{itemize}
In this work, we ask the following question: what local quantum field theory models can exhibit these transitions for generalized symmetries? 
The question is challenging partially because higher-form symmetries \cite{Gaiotto:2014kfa} or symmetries from condensation defects (e.g. \cite{Gaiotto:2019xmp,Roumpedakis:2022aik,Choi2023duality,Cordova:2024mqg}) do not act on local operators in quantum field theories, and thus it is challenging to use them to distinguish different renormalization group flows: any flow by local operators automatically preserves the symmetries and thus has the same anomalies. This is in contrast to lattice models, where local Hamiltonian terms are allowed to have finite nonzero radius of support.

There are previous attempts to construct symmetry breaking phase transitions of one-form symmetries using string-like non-local fields \cite{Iqbal:2021rkn} instead of local quantum fields. However, such theories are not readily accessible in experiments, since the field theories for most condensed matter systems are still described by local quantum fields such as  electrons.

In this work we will construct continuum field theory models with local quantum fields for these transitions for generalized symmetries in 3+1d and 2+1d. We will focus on one-form symmetries, and symmetries from their condensation defects. We will construct the models using gauge theories with bosons or fermions, where the phase transitions are induced by Higgs transitions, or fermion mass transitions where the masses flip sign. 
The gauge theory models have electric one-form symmetry and magnetic symmetry, which can exhibit SETs in the unbroken phases, which is crucial for matching the anomalies for these symmetries. 

The models discussed here can serve as building blocks to detect phase transitions in more complicated theories such as finite density QCD as discussed in \cite{Dumitrescu:2023hbe,QCD2025}. 
Since any local operator perturbations preserve higher-form symmetries or symmetries from condensation defects, we can deduce the phase transitions even in the presence of arbitrary potential for the local quantum fields as long as the asymptotic semiclassical phases remain the same. This is similar to the anomaly in the space of coupling or higher Berry phases in quantum field theories \cite{Cordova:2019jnf,Cordova:2019uob,Hsin:2020cgg}.

Throughout the discussion we will focus on invertible one-form symmetries. The SPT or SET of one-form symmetry also corresponds to SPT or SET for the symmetries generated by their condensation defects. Different SPTs and SETs correspond to different results under gauging the symmetries. Since condensation defects correspond to gauging (or proliferating) the one-form symmetry on submanifolds, gauging the symmetry of condensation defect corresponds to gauging the one-form symmetry on the entire space.
Thus SPTs or SETs for one-form symmetry corresponds to SPTs or SETs for the symmetry from the condensation defects.
We remark that the symmetries from condensation defects cannot be broken, since they admit gapped boundaries that allow them to unbraid with any potential order operators.

Systems with unbroken one-form symmetries can belong to nontrivial SPT or SET phases. Trivially gapped systems can be in different SPT phases with one-form symmetries
\cite{Gaiotto:2014kfa,Hsin:2018vcg,Tsui:2019ykk}.
Topologically ordered systems with unbroken one-form symmetry can be in different SET phases with different fractionalizations of one-form symmetry.
SETs of one-form symmetries in 3+1d have been discussed in e.g. \cite{Hsin:2019fhf,Hsin:2025jot}.
We will systematically study the fractionalization of one-form symmetry in general continuous group gauge theories in 3+1d. In particular, we show that continuous gauge theories can have different phases with the same unbroken one-form symmetry and the same topological order, but different fractionalizations of one-form symmetry. We also construct continuous gauge theories with phase transitions between broken one-form symmetry and unbroken fractionalized one-form symmetry, where both phases have the same $\mathbb{Z}_2$ topological order.

The work is organized as follows. In section \ref{sec:fractionalizationoneform}, we discuss the fractionalization of electric and magnetic one-form symmetries in 3+1d continuous gauge theories. In section \ref{sec:SSB}, we discuss models that exhibit symmetry breaking transitions. In section \ref{sec:SPT}, we  discuss models that exhibit SPT transitions. In section \ref{sec:SET}, we discuss models that exhibit SET transitions. We discuss future directions in section \ref{sec:outlook}.
In Appendix \ref{sec:thetaangle} we summarize several useful properties of theta angles.

\begin{figure}[t]
    \centering
    \includegraphics[width=0.3\linewidth]{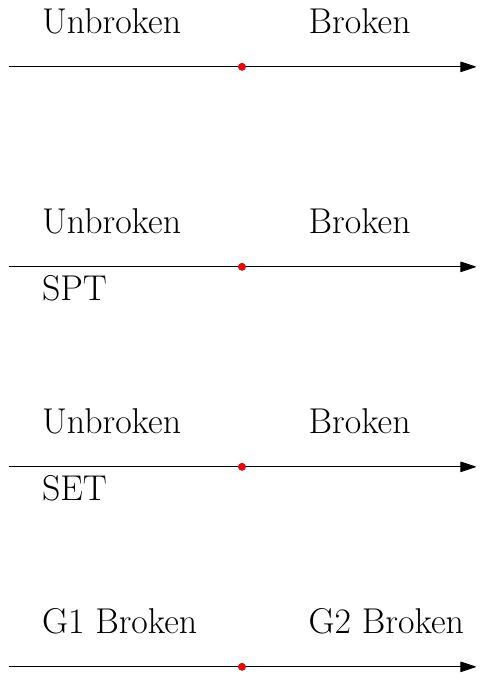}
    \caption{Scenarios for symmetry breaking transitions.}
    \label{fig:SSBsketch}
\end{figure}

\section{Fractionalization of Electric/Magnetic One-form Symmetries in 3+1d}
\label{sec:fractionalizationoneform}

Consider gauge theory in 3+1d with continuous connected gauge group $G$. Let us focus on $G$ such that it does not contain $U(1)$ factor. In this section we will discuss pure $G$ gauge theory with zero theta angle. Let us denote $G=\tilde G/C$ for universal covering $\tilde G$ and finite Abelian $C\subset Z(\tilde G)$. We will focus on cyclic $C=\mathbb{Z}_N$ for some integer $N$ in the following discussion; in the more general case this amounts to considering different cyclic subgroups of $C$.

The theory has $Z(G)=Z(\tilde G)/C$ center electric one-form symmetry that transforms Wilson lines linearly, as well as $\text{Hom}(\pi_1(G),U(1))\cong C$ magnetic one-form symmetry that transforms 't Hooft lines linearly. 
We will exam the fractionalization of these symmetries at low energy.

\subsection{Fractionalization of electric one-form symmetry}

At low energy, since $\tilde G$ gauge theory confines and is trivially gapped, the theory flows to two-form $C$ gauge theory, where the $C$ Wilson lines are the deconfined monopoles in the $G$ gauge theory that align with the monopole condensation. The $G$ Wilson lines are confined and thus the electric one-form symmetry is unbroken.  Let us exam the fractionalization of the electric one-form symmetry.

We will show that the generator of the broken magnetic one-form symmetry carry fractional charge under the electric one-form symmetry.
The generator of the magnetic symmetry can be expressed in terms of the Cartan component of the $G$ gauge field, denoted by $\{a_i\}$ for $i=1,\cdots,r$ where $r:=\text{rank}(G)$, as
\begin{equation}
    w_2^{G}=\sum_i \frac{da_i}{2\pi}\text{ mod }N~,
\end{equation}
where recall that the magnetic symmetry is $C=\mathbb{Z}_N$. This is the sum of GNO charges \cite{GODDARD19771} mod $N$. 
The generator of the magnetic one-form symmetry is
\begin{equation}
    U_m^\ell(\Sigma)=e^{\frac{2\pi i\ell}{N}\int_\Sigma w_2^G}~.
\end{equation}
Since the electric center one-form symmetry shifts the Cartan gauge fields, the low energy $C=\mathbb{Z}_N$ gauge theory has loops that carry fractional charge $\frac{r}{N}\mathbb{Z}$ under the electric one-form symmetry, which is a nontrivial fraction when $N\neq \gcd(N,\gcd(r,N))$, and the order of the electric one-form symmetry is not coprime with $N/\gcd(r,N)$.

\subsection{Fractionalization of magnetic one-form symmetry in Higgs phase}

Consider the Higgs phase where the $G$ gauge group is broken to a finite subgroup $K$. The electric one-form symmetry acts linearly on the $K$ Wilson lines that come from decomposition of $G$ Wilson lines with nonzero charge. If such $K$ Wilson lines exist then the electric one-form symmetry is broken, since the $K$ Wilson lines are deconfined in finite group gauge theory.
Moreover, since $K$ gauge field does not have nonzero local field strength, there is no monopole at low energy and
the magnetic one-form symmetry is unbroken. Let us exam the fractionalization of the magnetic one-form symmetry at low energy.

If we turn on background gauge field $B$ for the magnetic one-form symmetry, it couples as
\begin{equation}
   \frac{2\pi}{N}\int  w_2^G\cup B~,
\end{equation}
where $B=0,1,\cdots,N-1$.
In terms of Cartan component gauge fields, and change to the normalization $B\in \frac{2\pi}{N}\mathbb{Z}$ it is
\begin{equation}
   \frac{1}{2\pi}\sum_i \int  da_i B~.
\end{equation}
The magnetic one-form symmetry fractionalizes when $K$ can be embedded in the Cartan subgroup. For example, consider $K=\mathbb{Z}_M$ where the $\mathbb{Z}_M$ gauge field corresponds to
\begin{equation}
    a_1=a,\quad a_{i>1}=0~,
\end{equation}
for $\mathbb{Z}_M$ gauge field $a$ with holonomy $\frac{2\pi}{M}\mathbb{Z}$.
The coupling to $B$ gives the low energy theory
\begin{equation}\label{eqn:ZMgaugetheory}
    \frac{1}{2\pi}daB+\frac{M}{2\pi}adb~,
\end{equation}
where we introduce Lagrangian multiplier $b$ to constrain $a$ to be a $\mathbb{Z}_M$ gauge field, and it describes the vortex of the Higgs phase.
Then the operator $e^{i\int b}$ carries fractional charge $\frac{1}{M}$ of the $\mathbb{Z}_N$ magnetic one-form symmetry, which is a nontrivial fraction when $\gcd(N,M)\neq 1$. Such fractionalization of one-form symmetry is also discussed in \cite{Hsin:2019fhf,Hsin:2025jot}.

Note that the fractionalization of magnetic symmetry does not require electric one-form symmetry, unlike the fractionalization of electric one-form symmetry. 

\subsection{Fractionalization and symmetry junction}

The fractionalization of one-form symmetry can also be described in terms of the junction of symmetry generator similar to \cite{Barkeshli:2014cna}. For example, in (\ref{eqn:ZMgaugetheory}), the one-form symmetry is generated by the operator
\begin{equation}
    U_\alpha =e^{i\alpha \int da},\quad \alpha\in \frac{2\pi}{M}\mathbb{Z}.
\end{equation}
Let us denote $[\alpha]$ to the the restriction of angular parameter $\alpha$ to $[0,2\pi)$. Then the junction of symmetry generators with $[\alpha_1],[\alpha_2],[\alpha_1+\alpha_2]$ is the line operator \cite{Hsin:2025jot}
\begin{equation}    U_{[\alpha_1+\alpha_2]}^{-1}U_{[\alpha_2]}U_{[\alpha_1]}=e^{iq(\alpha_1,\alpha_2)\int a},\quad q(\alpha_1,\alpha_2)=\frac{[\alpha_1]+[\alpha_2]-[\alpha_1+\alpha_2]}{2\pi}\in\mathbb{Z}~.
\end{equation}
This fact that the junction is a nontrivial line operator implies that the surface operators that braid with the line operator has nontrivial fractional one-form charge.

\subsection{Fractionalization and anomaly matching}

The fractionalization of one-form symmetries is crucial for anomaly matching. The electric and magnetic one-form symmetries in the $G$ gauge theory has a mixed anomaly when $Z(\tilde G)$ is a non-split extension of $C$,
\begin{equation}
    1\rightarrow C\rightarrow Z(\tilde G)\rightarrow Z(\tilde G)/C~.
\end{equation}
To see the mixed anomaly, we note that gauging the $C$ magnetic one-form symmetry extends the gauge group from $G$ to $\tilde G$, and the electric one-form symmetry is extended from $Z(\tilde G)/C$ to $Z(\tilde G)$. If $Z(\tilde G)/C$ is not a subgroup of $Z(\tilde G)$, this means a mixed anomaly (e.g. \cite{Benini:2018reh,Hsin:2020nts}).

Such anomaly in the UV gauge theory should be matched in the low energy $C$ two-form gauge theory. The mixed anomaly comes from the fractional charge of the generator of the magnetic one-form symmetry under the electric one-form symmetry. Thus the fractionalization of one-form symmetry is crucial for anomaly matching. 

In particular, the electric one-form symmetry fractionalizes when there is magnetic one-form symmetry and mixed anomaly between the electric and magnetic one-form symmetries 
described by the bulk term of the form $\int B^e dB^m$ for backgrounds $B^e,B^m$ of the electric and magnetic one-form symmetries.

\paragraph{Example: $SO(2n)$ gauge theory with $n\geq 2$.}

Consider $G=SO(2n)$, $C=\mathbb{Z}_2$.
The theory has $\mathbb{Z}_2$ electric one-form symmetry and $\mathbb{Z}_2$ magnetic one-form symmetry. 
At low energy, the generator of the magnetic symmetry carries fractional charge $n/2$ under the electric one-form symmetry, which is an integer for even $n$ and a fraction for odd $n$. 

This is consistent with the property that the two $\mathbb{Z}_2$ one-form symmetries have a mixed anomaly when $n$ is odd, where the center of $\tilde G=Spin(2n)$ is $\mathbb{Z}_4$, and no mixed anomaly when $n$ is even, where the center of $\tilde G=Spin(2n)$ is $\mathbb{Z}_2\times\mathbb{Z}_2$ \cite{Cordova:2017vab,Benini:2018reh,Ang:2019txy,Hsin:2020nts}.

\paragraph{Example: $SU(mn)/\mathbb{Z}_n$ gauge theory}

Consider $G=SU(mn)/\mathbb{Z}_n$, $C=\mathbb{Z}_n$.
The theory has $\mathbb{Z}_m$ electric one-form symmetry and $\mathbb{Z}_n$ magnetic one-form symmetry. 
The symmetries have mixed anomaly when the extension $1\rightarrow\mathbb{Z}_n\rightarrow\mathbb{Z}_{mn}\rightarrow\mathbb{Z}_m\rightarrow1$ does not split, i.e. when $\gcd(m,n)\neq 1$. Indeed, the fractional charge is nontrivial when the order of electric one-form symmetry, $m$, is not coprime with $n$ \cite{Benini:2018reh,Ang:2019txy,Hsin:2020nts}.

\subsection{Comment on induced fractionalization of 0-form symmetries}

Let us briefly comment that fractionalization of one-form symmetry can also induce a fractionalization of 0-form symmetries
via the relation that express the backgrounds of the one-form symmetries in terms of backgrounds of 0-form symmetries \cite{Benini:2018reh,Hsin:2019fhf}. We will call it induced fractionalizations. 
For example, in the $\mathbb{Z}_M$ gauge theory (\ref{eqn:ZMgaugetheory}) the fractionalization of the magnetic one-form symmetry can be described by the relation between background $B_3$ for two-form symmetry generated by $e^{i\int a}$ and $B$ of one-form symmetry as $B_3=\frac{1}{M} dB$. Then the relation between $B$ and background of 0-form symmetry induces a relation between $B_3$ and the 0-form symmetry background, which describes a fractionalization of 0-form symmetry \cite{Hsin:2019fhf}.

We will illustrate this mechanism in detail in a separate work, including continuous gauge theory models in 3+1d for fermionic loop excitations \cite{Thorngren:2014pza,Fidkowski:2021unr,Chen:2021xks,Kobayashi:2024dqj} and mixed anomalies of one-form symmetries and gravity
from Lorentz symmetry fractionalization \cite{Hsin:2019gvb}.

\section{Generalized Symmetry Breaking Transitions}
\label{sec:SSB}

We will discuss local quantum field theory models that exhibit symmetry breaking phase transitions for exact higher-form symmetries or symmetries from condensation defects. The models are gauge theories with bosons or fermions. Across the phase transitions, the symmetries change from spontaneously broken to unbroken, or different symmetries are broken, similar to deconfined quantum criticalities. When the symmetries are unbroken, there can be SPT phases or fractionalizations. We sketch the scenarios in Fig.~\ref{fig:SSBsketch}.

\subsection{Gauge theories with bosons}

Consider $G$ gauge theory with Higgs scalar that can break the gauge group to subgroup $K$ gauge theory. Here, $G=\tilde G/C$ is a continuous connected group with universal covering $\tilde G$ and $C\subset Z(\tilde G)$. In 3+1d, the theory has the following one-form symmetries.

\paragraph{Electric center one-form symmetry.}
If the Higgs fields are invariant under $Z'\subset Z(G)$, the theory has center one-form symmetry $Z'$. The center one-form symmetry acts linearly on the Wilson lines in $G$ and $K$ gauge theories.

When the Higgs fields do not condense, the theory flows to pure $G$ gauge theory. For $G$ containing $U(1)$ factors the electric one-form symmetry is spontaneously broken by the $U(1)$ Wilson lines. Let us focus on the cases where $G$ does not contain $U(1)$ factors. Then the Wilson lines confine and the electric one-form symmetry is unbroken. 
Whether the unbroken phase is a nontrivial SPT or SET depends on whether there is topological terms for $G$ gauge field and whether it is simply connected. If $G$ is simply connected, then the theory is trivially gapped, and the SPT phase of the center one-form symmetry depends on the theta angles of $G$ \cite{Gaiotto:2014kfa,Hsin:2018vcg,Hsin:2020nts}. If $G$ is not simply connected and it does not contain $U(1)$ factor, the theory flows to TQFT given by two-form $C$ gauge theory. The particles in the two-form gauge theory correspond to the monopoles in the UV $G$ gauge theory, which do not transform under the electric one-form symmetry, and thus the center one-form symmetry is always unbroken.
Depending on the theta angle, the two-form $C$ gauge theory can have deconfined loop excitations, and the loops can carry fractional charge under the center one-form symmetry and gives nontrivial SET as discussed in section \ref{sec:fractionalizationoneform}.

In the Higgs phase, the fate of the symmetry depends on $K$:
\begin{itemize}
    \item If $K$ is a discrete group, the field strengths are locally trivial and the $K$ Wilson lines are deconfined: the electric one-form symmetry is spontaneously broken in the Higgs phase. Thus the Higgs transition drives one-form symmetry breaking. 

    \item 
If $K$ is also a continuous connected group, the Wilson lines are confined, and the electric one-form symmetry is also unbroken in the Higgs phase.

\end{itemize}

\paragraph{Magnetic one-form symmetry.}
 The magnetic one-form symmetry $\text{Hom}(\pi_1(G),U(1))=\text{Hom}(C,U(1))$.

When the Higgs fields do not condense, the fate the of the symmetry depends on the topological terms of $G$ and whether there is $U(1)$ factor in $G$. When there is $U(1)$ factor, the low energy theory contains free $U(1)$ gauge theory and the $U(1)$ par of the magnetic symmetry is broken regardless of theta angles. Let us focus on the case that $G$ does not contain $U(1)$ factors. Depending on the theta angles, the monopoles can be confined which make the corresponding magnetic symmetry unbroken \cite{Gaiotto:2014kfa,Hsin:2018vcg}. Such unbroken phase can have nontrivial SPTs to be discussed in section \ref{sec:SPT}.
For zero theta angles all magnetic symmetry is spontaneously broken in the phase where the Higgs fields do not condense.

When the Higgs fields condense, the fate of the symmetry depends on $K$ and topological terms. For simplicity, let us set theta angles of $G$ to zero. Some of the monopoles confine in the Higgs phase due to the electric charge condensation, and the corresponding unbroken magnetic symmetry is the subgroup
\begin{equation}
    \text{Unbroken magnetic symmetry:}\quad \text{Ker }\iota'^*~,
\end{equation}
where $\iota$ is the inclusion $\iota:K\rightarrow G$, and it lifts to $\iota':\pi_1(K)\rightarrow \pi_1(G)$. The pullback gives the map $\iota'^*:\text{Hom}(\pi_1(G),U(1))\rightarrow \text{Hom}(\pi_1(K))$. The kernel corresponds the magnetic symmetries that act trivially on the monopoles of the $K$ gauge theory and thus unbroken. Furthermore, in the Higgs phase for disconnected group $K$, vortex loops can carry fractional charge under the magnetic symmetry and gives nontrivial SET as discussed in section \ref{sec:fractionalizationoneform}.

Throughout the discussions with Higgs fields, we will introduce suitable perturbations to lift any Goldstone modes.

\subsubsection{SSB-SET transition: $U(1)$ gauge theory}
\label{sec:U1example}

\begin{figure}[t]
    \centering
    \includegraphics[width=0.8\linewidth]{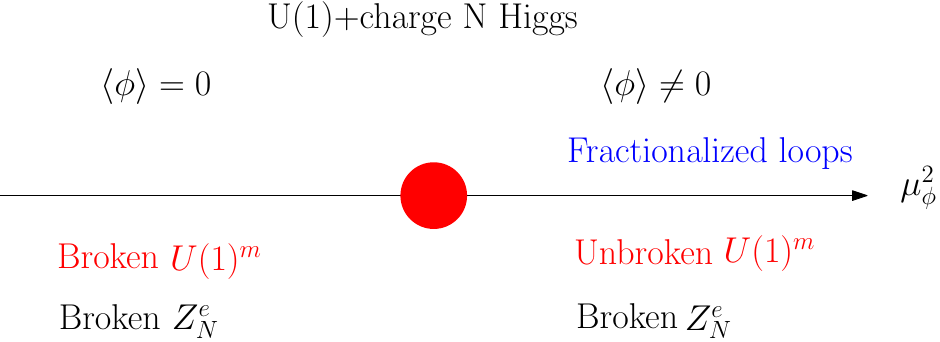}
    \caption{Phase diagram of the Higgs transition for $U(1)$ gauge theory with charge $N$ scalar in 3+1d, where the magnetic symmetry is a one-form symmetry. The Higgs potential is $V(\phi)=-\mu^2_\phi|\phi|^2+\lambda|\phi|^4$ with $\lambda>0$.}
    \label{fig:u(1)+chargeN}
\end{figure}

Consider $U(1)$ gauge theory with Higgs field of charge $N$ in $D$ spacetime dimension. The theory has $U(1)^{(D-3)}_m$ $(D-3)$-form magnetic symmetry as well as $\mathbb{Z}_N$ electric one-form center symmetry, where the $U(1)$ center one-form symmetry is explicitly broken to $\mathbb{Z}_N$ since the Wilson line of charge $n$ can end on the Higgs field.

Let us analyze the phases depending on whether the Higgs field $\phi$ condenses;
\begin{itemize}
    \item When the Higgs scalar does not condense $\langle\phi\rangle =0$, we get free $U(1)$ gauge theory, with spontaneously broken $U(1)_m$ symmetry. The electric one-form symmetry is enlarged from $\mathbb{Z}_N$ to emergent $U(1)_e$ and is also spontaneously broken.

\item When the Higgs scalar condenses $\langle\phi\rangle =v$, the gauge group is broken to $\mathbb{Z}_N$. The $\mathbb{Z}_N$ electric one-form symmetry is spontaneously broken by the deconfined $\mathbb{Z}_N$ Wilson lines. 
On the other hand, since $\mathbb{Z}_n$ gauge field does not have fluxes on sphere, the magnetic monopoles confine, and the $U(1)_m$ magnetic symmetry is unbroken. 

\item
While the magnetic symmetry is unbroken in the Higgs phase, the symmetry still acts non-trivially at low energy. As shown in \cite{Hsin:2019fhf}, the magnetic symmetry acts by fractionalization. Let us review the argument below. if we turn on background field $B$ for the magnetic symmetry, it couples to the dynamical $U(1)$ gauge field $a$ as
\begin{equation}
    \frac{da}{2\pi} B~.
\end{equation}
In the Higgs phase, we can dualize the phase of the Higgs field to $(D-2)$-form $b$, then the low energy theory is
\begin{equation}
    \frac{N}{2\pi}da b + \frac{1}{2\pi}da B~.
\end{equation}
The equation of motion of $a$ then implies that the vortex operator $e^{i\int b}$ carries fractional magnetic charge $1/N$.
In 3+1d, the fractional excitations are fractional loops that carry fractional charge $1/N$ under the magnetic one-form symmetry. They are also called ANO vortices, which carry fractional magnetic fluxes that are not multiple of $2\pi$, i.e. ``a fraction of monopoles''. Similar fractionalizations for higher-form symmetries are also studied in \cite{Hsin:2025jot}.

\end{itemize}
The above Higgs transition describes the confinement transition of monopoles whose magnetic charge is not a multiple of $N$. This corresponds to the symmetry breaking transition of the magnetic symmetry. The electric one-form symmetry is spontaneously broken in the two phases.
The phase diagram is sketched in Fig.~\ref{fig:u(1)+chargeN}.
We note that in 3+1d, the Coleman-Weinberg quantum correction \cite{PhysRevD.7.1888} is expected to render the transition first order.

\paragraph{Anomaly matching in Higgs phase}
Let us demonstrate the fractionalization of magnetic one-form symmetry is necessary for matching the mixed anomaly between the electric one-form symmetry and the magnetic one-form symmetry.
The $\mathbb{Z}_N$ electric one-form symmetry and the magnetic $U(1)$ one-form symmetry has a mixed anomaly described by the 4+1d bulk term \cite{Gaiotto:2014kfa}
\begin{equation}
    \frac{1}{2\pi}B' dB~,
\end{equation}
where $B'$ is the background for the electric one-form symmetry.
The anomaly is matched in the Higgs phase: the action with both $B,B'$ turned on is
\begin{equation}\label{eqn:ZN+Be+Bm}
    \frac{N}{2\pi}dab+\frac{1}{2\pi}daB+\frac{N}{2\pi}B'b~.
\end{equation}
Under the background gauge transformation
\begin{equation}
    B\rightarrow B+d\lambda,\quad B'\rightarrow B'+d\lambda',\quad a\rightarrow a-\lambda'~,
\end{equation}
the action (\ref{eqn:ZN+Be+Bm}) transforms by the anomalous term $\left(-\frac{1}{2\pi}d\lambda' B\right)$. This is compensates by the inflow from the bulk term:
\begin{equation}
\frac{1}{2\pi}\int B' dB\;\rightarrow \;\frac{1}{2\pi}\int B' dB+\frac{1}{2\pi}\int_\text{bdy}d\lambda' B~.
\end{equation}

\subsubsection{SSB-SET transition: $SO(2n+1)$ gauge theory}

\begin{figure}[t]
    \centering
    \includegraphics[width=0.5\linewidth]{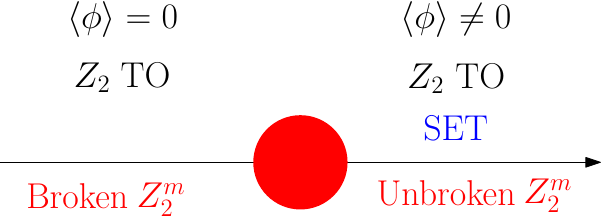}
    \caption{Phase diagram of $SO(2n+1)$ gauge theory with Higgs fields that can break the gauge group to $\mathbb{Z}_2$ in 3+1d. Both phases are the same $\mathbb{Z}_2$ gauge theory topological order, but the $\mathbb{Z}_2$ magnetic one-form symmetry acts differently in the two phases.
    The magentic one-form symmetry fractionalizes in the unbroken phase, which is a nontrivial SET. }
    \label{fig:SSBSETSO2n+1}
\end{figure}

Consider $SO(2n+1)$ gauge theory with Higgs fields that can break the gauge group to a $\mathbb{Z}_2$ subgroup.\footnote{
For example, one can add several vector Higgs that have vev $(1,0\cdots,0)$, $\cdots$, $(0,0,\cdots,1,0,0)$, and an adjoint Higgs with vev $\text{diag}(0,\cdots,0,1)$.
}
The theory has $\mathbb{Z}_2$ magnetic one-form symmetry. 

When the Higgs fields do not condense, the theory flows to $\mathbb{Z}_2$ gauge theory topological order with broken magentic one-form symmetry.

When the Higgs fields condense, the theory flows to $\mathbb{Z}_2$ gauge theory with deconfined Wilson line and broken emergent electric one-form symmetry. The magnetic one-form symmetry is still unbroken, but it has nontrivial fractionalization in the $\mathbb{Z}_2$ gauge theory topological order.

In conclusion, the Higgs transition is between two phases with the same $\mathbb{Z}_2$ topological order, but the $\mathbb{Z}_2$ magnetic one-form symmetry is broken in one phase, and unbroken in the other phase with nontrivial fractionalization of the one-form symmetry.

The phase diagram is sketched in Fig.~\ref{fig:SSBSETSO2n+1}.

\subsubsection{Analogue of DQCP from $SO(2n)$ and $SU(N_1N_2)/\mathbb{Z}_{N_2}$ gauge theories}

\begin{figure}[t]
    \centering
    \includegraphics[width=0.55\linewidth]{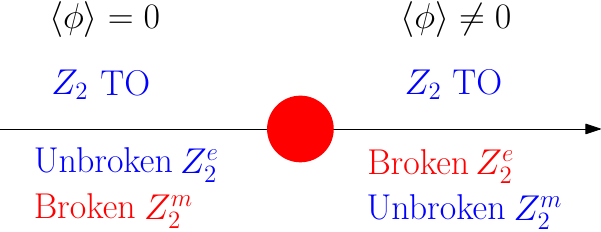}
    \caption{Phase diagram for $SO(2n)$ gauge theory with Higgs fields in 3+1d. This is an analogue of DQCP for $\mathbb{Z}_2^e\times\mathbb{Z}_2^m$ one-form symmetry in 3+1d. When $n$ is odd, the two phases have nontrivial SETs.}
    \label{fig:DQCP}
\end{figure}

In the original deconfined quantum critical phase (DQCP) \cite{Senthil_2004}, the N\'eel phase spontaneously breaks the $SO(3)$ spin rotation symmetry, while the VBS phase spontaneously breaks the $O(2)$ rotation symmetry (on the lattice, only a discrete rotation subgroup is manifest).
Here, we will generalize the DQCP in the following sense: the phase transition is separated by spontaneously symmetry breaking of two different ``sub-symmetries'' of the full symmetry. Examples of such generalization are discussed in e.g. \cite{Zhang_2023}.

\paragraph{$SO(2n)$ gauge theory}

Consider $SO(2n)$ gauge group (with $n\geq 2$) broken to $\mathbb{Z}_2$ gauge group in the center of $SO(2n)$ by adjoint Higgs fields. The theory has $\mathbb{Z}_2$ center electric one-form symmetry and $\mathbb{Z}_2$ magnetic one-form symmetry.
Let us take the theta angle to zero.

When the Higgs fields do not condense, the Wilson lines confine and the monopole is deconfined, leading to unbroken electric one-form symmetry and broken magnetic one-form symmetry. The low energy is $\mathbb{Z}_2$ two-form gauge theory. When $n$ is even, there is no mixed anomaly between the electric and magentic one-form symmetry, and thus the electric one-form symmetry does not fractionalize. 
On the other hand, when $n$ is odd, there is mixed anomaly between the electric and magnetic one-form symmetry \cite{Benini:2018reh,Hsin:2020nts}, and the unbroken electric one-form symmetry fractionalizes in the low energy $\mathbb{Z}_2$ topological order: the loop excitation carry charge $1/2$ under the electric one-form symmetry to match the anomaly. If the theta angle is nonzero, there can be additional SPT phase.

When the Higgs fields condense, the gauge group is broken to the center $\mathbb{Z}_2$. The Wilson line becomes deconfined and the electric one-form symmetry is spontaneously broken. On the other hand, the monopole confines and the magnetic one-form symmetry is unbroken. The magnetic one-form symmetry fractionalizes on the loop excitation when $n$ is odd and no fractionalization when $n$ is even.

In conclusion, both the uncondensed and condensed phases have $\mathbb{Z}_2$ topological orders, but different subgroups of $\mathbb{Z}_2\times\mathbb{Z}_2$ electric and magnetic one-form symmetry is spontaneously broken in the two phases. 
We sketch the phase diagram for zero theta angle in Fig.~\ref{fig:DQCP}.
We remark that this is similar to the deconfined quantum criticalities (DQCP) where different subgroups of $SO(3)\times U(1)$ is broken across the transition  (e.g. \cite{Senthil:2023vqd}).

\begin{figure}[t]
    \centering
    \includegraphics[width=0.5\linewidth]{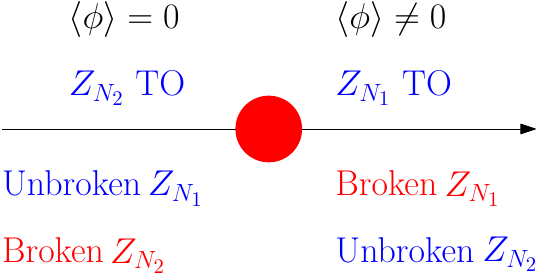}
    \caption{Phase diagram of $SU(N_1N_2)/\mathbb{Z}_{N_2}$ gauge theory with adjoint Higgs fields in 3+1d. This is an analogue of DQCP for $\mathbb{Z}_{N_1}^e\times\mathbb{Z}_{N_2}^m$ one-form symmetry in 3+1d. When $\gcd(N_1,N_2)\neq 1$, the two phases have nontrivial SETs.}
    \label{fig:DQCP2}
\end{figure}

\paragraph{$SU(N_1N_2)/\mathbb{Z}_{N_2}$ gauge theory}

We can generalize the model to $SU(N_1N_2)/\mathbb{Z}_{N_2}$ gauge theory with adjoint Higgs that has $\mathbb{Z}_{N_1}^e$ electric one-form symmetry and $\mathbb{Z}_{N_2}^m$ magnetic one-form symmetry. Then the topological orders in the uncondensed and condensed phases will be $\mathbb{Z}_{N_2},\mathbb{Z}_{N_1}$ gauge theories, corresponding to broken magnetic one-form symmetry and broken electric one-form symmetry, respectively, and the unbroken one-form symmetries fractionalize when $\gcd(N_1,N_2)\neq 1$. We sketch the phase diagram in Fig.~\ref{fig:DQCP2}.

\subsubsection{SSB-SPT transition: $SU(N)$ gauge theory}

\begin{figure}[t]
    \centering
    \includegraphics[width=0.6\linewidth]{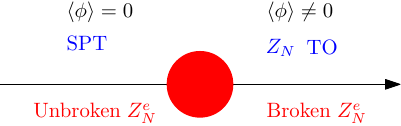}
    \caption{Phase diagram of $SU(N)$ gauge theory with adjoint Higgs fields in 3+1d. This is a symmetry breaking transition for $\mathbb{Z}_{N}^e$ electric center one-form symmetry.}
    \label{fig:SSBSPTboson}
\end{figure}

Consider $SU(N)$ gauge theory with adjoint Higgs fields that break the gauge group to $\mathbb{Z}_N$. The theory has $\mathbb{Z}_N$ electric one-form symmetry. We can turn on theta angle $\theta=2\pi k$.

When the Higgs fields do not condense, the theory flows to $SU(N)$ pure gauge theory, which is confined and trivially gapped. The one-form symmetry is unbroken. If the theta angle is $2\pi k$, the phase has SPT for the electric one-form symmetry \cite{Hsin:2018vcg}
\begin{equation}
    2\pi\frac{k(N-1)}{2N}\int {\cal P}(B)~,
\end{equation}
where $B$ is the background two-form gauge field for the one-form symmetry. Note that this is equivalent to tensoring the theory with zero theta angle and the SPT.

When the Higgs fields condense and break the gauge group to $\mathbb{Z}_N$ center, the electric one-form symmetry is spontaneously broken regardless of the theta angles due to the deconfined $\mathbb{Z}_N$ Wilson lines. We sketch the phase diagram in Fig.~\ref{fig:SSBSPTboson}. 

\subsubsection{More example: $SO(2n)$ gauge theory}

Consider $SO(2n)$ gauge group broken to $SU(n)$ gauge group, with $n\geq 2$. The theory has $\mathbb{Z}_2$ magnetic one-form symmetry. When $n$ is even, the theory has $\mathbb{Z}_2$ center electric one-form symmetry. The electric one-form symmetry is unbroken in the two phases, and we will focus on magnetic one-form symmetry. 

\paragraph{Zero theta angle}
When the Higgs fields do not condense, the $SO(2n)=Spin(2n)/\mathbb{Z}_2$ gauge theory flows to $\mathbb{Z}_2$ two-form gauge theory due to confinement of $Spin(2n)$, and the magnetic symmetry is spontaneously broken for zero theta angle.

When the Higgs fields condense and break the gauge group to $SU(n)$, there is no monopoles charged under the magnetic symmetry, and the magnetic symmetry is unbroken.
In the Higgs phase, the theory is trivially gapped: the $SU(n)$ gauge theory confines. Thus there is no fractional excitations in Higgs phase. 
The phase diagram is the standard Landau paradigm diagram with $\mathbb{Z}_2$ topological order in the uncondensed phase.

\paragraph{Nonzero theta angle}

Let us take $n\neq 2$ to be even. Consider $SO(2n)$ gauge theory with $\theta=2\pi$. The theory has $\mathbb{Z}_2$ electric center one-form symmetry and $\mathbb{Z}_2$ magnetic one-form symmetry. The theta angle $\theta=2\pi$ for $n\geq 2$ is equivalent to $\theta=0$ and the discrete theta angle
\begin{equation}
    \pi\int w_2^{SO}\cup w_2~.
\end{equation}

In the Higgs phase, this also gives $\theta=2\pi$ for $SU(n)$.\footnote{
To see this, note that 2+1d $2N$ massless Majorana fermions in the vector representation of $SO(2n)$ is equivalent to $N$ massless Dirac fermions in fundamental representation $SU(n)$. The fermion mass transition gives the jump of topological terms $SO(2n)_1$ and $SU(n)_1$ respectively.
}
The theory is trivially gapped and both one-form symmetries are unbroken. The $\mathbb{Z}_2$ center one-form symmetry has nontrivial SPT due to $\theta=2\pi$: 
\begin{equation}
    2\pi\frac{n-1}{2n}(n/2)^2\int {\cal P}(B^e)=\frac{2\pi}{4}\frac{n(n-1)}{2}\int {\cal P}(B^e)
    ~,
\end{equation}
where $B^e$ is the background for electric one-form symmetry. This equals $\pm\frac{2\pi}{4}\int {\cal P}(B^e)$ for $n=2,6$ mod 8, trivial for $n=0$ mod 8 and $\pi\int B^e\cup B^e$ for $n=4$ mod 8. 

When the Higgs fields do not condense, 
 the low energy theory is $\mathbb{Z}_2$ gauge theory with fermion charge. The low energy theory in the presence of backgrounds $B^e,B^m$ for the electric and magnetic one-form symmetry can be obtained from (\ref{eqn:discthetaSO4n}) (where $n$ there is replaced by $n/2$)
 \begin{equation}
     \pi\int b\cup w_2+\pi\int b\cup (B^e+B^m)+\frac{2\pi (n/2)}{4}\int {\cal P}(B^e)~,
 \end{equation}
 where $b$ is a dynamical two-form gauge field that matches with $w_2^{SO}$. Thus the SPT phase of $B^e$ does not change across the transition. In addition, the diagonal $\mathbb{Z}_2$ one-form symmetry that corresponds to $B^e=B^m=B$ is unbroken, this is due to the property that monopoles acquire electric charges from the theta angle \cite{Witten:1979ey,Aharony:2013hda,Barkeshli:2022edm}. The other one-form symmetry is spontaneously broken.
Thus the Higgs transition is a symmetry breaking transition.

\subsection{Gauge theories with fermions}

Let us consider transitions where fermion mass change from positive to negative. From the fermion path integral and APS index theorems \cite{Atiyah:1975jf,Witten:2015aba}, this induces additional topological terms. In 3+1d gauge theories with continuous and connected gauge group, the topological terms are theta terms.

Take the gauge group $G$ to be a continuous Lie group, and $G=\tilde G/C$ for universal covering $\tilde G$ and finite Abelian $C\subset Z(\tilde G)$ in the center. 
The theory has magnetic symmetry $\text{Hom}(C,U(1))$. 
The gauge field couples to $N_f$ flavors of Dirac fermions in irreducible representation $R$ of the gauge group. By changing the fermion masses from positive to negative, the theta angle is shifted by
\begin{equation}
    \Delta \theta =  2\pi x_R N_f~,
\end{equation}
where $x_R$ is the Dynkin index or half of the Dynkin index for complex or real (and pseudo real) representation $R$. The shift can be understood as the fractional Hall response of massless Dirac fermion in 2+1d on the domain wall that interpolates between the positive mass and negative mass. 
When the number of flavor $N_f$ of Dirac fermion is odd, the theta angle can increases by $\pi$, which can lead to spontaneously breaking of (emergent) time-reversal symmetry  \cite{Gaiotto:2017yup}.
We will focus on the case of even number of flavor $N_f\in2\mathbb{Z}$ of Dirac fermions, then the shift of theta angle is always integer multiple of $2\pi$:
\begin{equation}
    \Delta \theta=2\pi k,\quad k=x_RN_f\in\mathbb{Z}~.
\end{equation}
Such shift of theta angle is equivalent to changing discrete theta angles, see Appendix \ref{sec:thetaangle}.

We will consider the model with zero theta angle for positive fermion masses, and the negative fermion masses produce $\theta=2\pi k$. 

For positive fermion masses, the magnetic one-form symmetry is completely broken due to deconfined monopoles that align with the monopole condensate, and the electric one-form symmetry is unbroken but can fractionalize as discussed in section \ref{sec:fractionalizationoneform}.

In the following, we will exam what happens to the one-form symmetries for negative fermion masses.

\subsubsection{SSB-SPT transition with trivial electric one-form symmetry}

 We will discuss the examples $PSO(4n+2)$, $PSO(4n)$, $PSp(N)$ separately. These cases do not have electric one-form symmetries, and the one-form symmetries do not fractionalize.
The low energy theories can be obtained from the discrete theta angles for $\theta=2\pi k$ in Appendix \ref{sec:thetaangle}.

\paragraph{Gauge group $G=PSO(4n+2)$.}
For $G=PSO(4n+2)$, the magnetic one-form symmetry is $\mathbb{Z}_4$. For example, adjoint Weyl fermion can realize the jump $k=2n$. We will focus on the cases of even $k$.

The low energy theory is
\begin{equation}
    2\pi \frac{k(2n+1)}{8}\int {\cal P}(b)+\frac{2\pi}{4}\int b\cup B^m~,
\end{equation}
where $b$ is a dynamical $\mathbb{Z}_4$ two-form gauge field that matches to $w_2^{PSO}$, and $B^m$ is the background for the one-form symmetry. Since $k$ is even, the theory has time-reversal symmetry.

The equation of motion for $b$ gives $k(2n+1)b+B^m=0$. Thus for $k=0$ mod 4, the magnetic one-form symmetry is fully broken. For $k=2$ mod 4, by taking mod 2 in the relation we find the magnetic one-form symmetry is broken to $\mathbb{Z}_2$.

When $k=2$ mod 4, let us express $b=2b'+b''$ for $\mathbb{Z}_2$ two-form gauge field $b,b'$. The unbroken one-form symmetry corresponds to the background $B^m=2B$.
Substituting it into the low energy action gives
\begin{equation}
    (k/2) (2n+1)  \frac{2\pi}{4}\int {\cal P}(b'')+ \pi\int b''\cup B~,
\end{equation}
and integrating out $b''$ gives the SPT
\begin{equation}
    -(k/2)(2n+1)\frac{2\pi}{4}\int {\cal P}(B)~,
\end{equation}
up to a gravitational term. Since $k=2$ mod 4, this is a nontrivial SPT.

Thus the fermion mass transition is a symmetry breaking transition for $k=2$ mod 4, where the $\mathbb{Z}_2$ subgroup one-form symmetry is broken/unbroken across the transition. The unbroken one-form symmetry has nontrivial SPT for the one-form symmetry.

\paragraph{Gauge group $G=PSO(4n)$.}

For $G=PSO(4n)$,  the magnetic one-form symmetry is $\mathbb{Z}_2\times\mathbb{Z}_2$. Adjoint Weyl fermion can realize the jump $k=2n-1$, and by taking multiple of them we need to discuss the cases of any integer $k$. The theory has time-reversal symmetry.

The low energy theory is
\begin{align}
    &\pi k\int b^{(1)}\cup b^{(1)}+\pi k\int b^{(1)}\cup b^{(2)}+\frac{2\pi nk}{4}\int {\cal P}(b^{(2)})+\pi\int b^{(1)}\cup B^{(1)}+\pi\int b^{(2)}\cup B^{(2)}\cr 
    &=\pi k\int b^{(1)}\cup \left(b^{(2)}+w_2\right)+\frac{2\pi nk}{4}\int {\cal P}(b^{(2)})+\pi\int b^{(1)}\cup B^{(1)}+\pi\int b^{(2)}\cup B^{(2)}~,
\end{align}
where $b^{(i)}$ are dynamical two-form gauge fields corresponding to $w_2^{(i)}$, and $B^{(i)}$ are the backgrounds for the $\mathbb{Z}_2\times \mathbb{Z}_2$ magentic one-form symmetry. In the second line we have used the Wu formula \cite{milnor1974characteristic}.
    \begin{itemize}
        \item     When $k=0$ mod 4, the above action is trivial and the low energy theory is $\mathbb{Z}_2\times\mathbb{Z}_2$ gauge theory with boson particles, and the magnetic one-form symmetry is fully broken.

\item     
    When $k=2$ mod 4, the first term in the second line is trivial. 
    
    If $n$ is even, the theory is $\mathbb{Z}_2\times\mathbb{Z}_2$ gauge theory with boson particles. If $n$ is odd, the theory is the
    tensor product of $\mathbb{Z}_2$ gauge theory with boson particle measured by $\int b^{(1)}$ and $\mathbb{Z}_2$ gauge theory with fermion particle measured by $\int b^{(2)}$.
    
    The magnetic one-form symmetry is fully broken.

\item     When $k=1,3$ mod 4, the low energy theory is trivially gapped due to the mixed term $b^{(1)}\cup b^{(2)}$, and the magnetic one-form symmetry is fully unbroken. Let us compute the SPT for the one-form symmetry.

Integrating out $b^{(1)}$ gives $b^{(2)}=B^{(1)}+w_2$, which results in the SPT
\begin{align}
    &\pi\int B^{(1)}\cup B^{(2)}+\pi\int B^{(2)}\cup B^{(2)}+
    \frac{2\pi nk}{4}\int {\cal P}(B^{(1)}+w_2)\cr
    &=\pi\int \left(B^{(1)}\cup B^{(2)}+ B^{(2)}\cup B^{(2)}\right)-\frac{2\pi nk}{4}\int {\cal P}(B^{(1)})+\frac{2\pi nk}{4}\int {\cal P}(w_2)~.
\end{align}

    \end{itemize}

    In conclusion, the $\mathbb{Z}_2\times\mathbb{Z}_2$ magnetic symmetry is completely broken for $k=0,2$ mod 4, and completely unbroken for $k=1,3$ mod 4.   
For $k=1,3$ mod 4 the fermion mass transition gives symmetry breaking transition with nontrivial SPT in the unbroken phase.
When $nk$ is even, the SPT is a cluster state of the $\mathbb{Z}_2\times\mathbb{Z}_2$ one-form symmetry.

\paragraph{Gauge group $G=PSp(N)$.}

For $G=PSp(N)$, the magnetic one-form symmetry is $\mathbb{Z}_2$.
The adjoint Weyl fermion can realize the jump $k=(N+1)/2$, and by taking multiple of them we need to discuss the cases with  any integer $k$.

The low energy theory is
\begin{equation}
2\pi \frac{Nk}{4}\int {\cal P}(b)+\pi\int b\cup B~,
\end{equation}
where $b$ is dynamical $\mathbb{Z}_2$ two-form gauge field corresponding to $w_2^{PSp}$, and $B$ is the background for  magnetic one-form symmetry.

When $Nk=0,2$ mod 4, the magnetic one-form symmetry is broken. The low energy theory is $\mathbb{Z}_2$ gauge theory with boson for $Nk=0$ mod 4 and $\mathbb{Z}_2$ gauge theory with fermion for $Nk=2$ mod 4.

When $Nk=1,3$ mod 4, the magnetic one-form symmetry is unbroken. Integrating out $b$ gives the SPT
\begin{equation}
    -2\pi \frac{Nk}{4}\int {\cal P}(B)~,
\end{equation}
up to a gravitational term. 
Thus the fermion mass transition realizes symmetry breaking transitions when $Nk=1,3$ mod 4, where the unbroken phase is nontrivial SPT of the one-form symmetry.

\paragraph{Example: $SO(3)$ gauge theory with adjoint fermion}

Consider $SO(3)$ gauge theory with adjoint fermion. The theory has $\mathbb{Z}_2$ magnetic one-form symmetry.

When the fermion mass is positive, this gives zero theta angle. The theory flows to $\mathbb{Z}_2$ gauge theory with boson particle, and the magnetic one-form symmetry is broken.

When the fermion mass is negative, the low energy theory is
\begin{equation}
    \frac{2\pi}{4}\int {\cal P}(b)+\pi\int b\cup B~,
\end{equation}
where $b$ is the dynamical two-form gauge field corresponding to $w_2^{SO}$.
The theory is trivially gapped, and the magnetic one-form symmetry is unbroken.
Integrating out $b$ gives the SPT
\begin{equation}
    -\frac{2\pi}{4}\int {\cal P}(B)~,
\end{equation}
up to a gravitational term. Thus the fermion mass transition realizes symmetry breaking transition of the magnetic one-form symmetry, with nontrivial SPT of the one-form symmetry in the unbroken phase.

\subsubsection{Analogue of DQCP from $SO(2n)$ gauge theory}

Consider $SO(2n)$ gauge theory for $n\geq 2$ with adjoint fermion. The theory has $\mathbb{Z}_2$ magnetic one-form symmetry and $\mathbb{Z}_2$ electric one-form symmetry. 

When the fermion mass is positive, the theta angle is zero, and the low energy theory is $\mathbb{Z}_2$ gauge theory with boson, where the magnetic one-form symmetry is broken while the electric one-form symmetry is unbroken. The electric one-form symmetry fractionalizes for odd $n$ and does not fractionalize for even $n$.

When the fermion mass is negative, the theta angle is $2\pi (n-2)$. The low energy theory can be obtained from (\ref{eqn:discthetaSO4n+2}),(\ref{eqn:discthetaSO4n}). Denote the backgrounds for the electric and magnetic one-form symmetries by $B^e,B^m$:
\begin{itemize}
    \item If $n$ is odd, we substitute $w_2^{PSO}=2b+B^e$ where the low energy dynamical field $b$ corresponds to $w_2^{SO}$ at low energy. The low energy theory is
    \begin{equation}
        2\pi \frac{(n-2)n}{8}\int {\cal P}(2b+B^e)+\pi\int b\cup B^m~.
    \end{equation}
The action contains $\pi\int b\cup b$, and thus
the low energy theory is the topological order of $\mathbb{Z}_2$ gauge theory with fermion charge. The action contains $\pi\int b\cup (B^e+B^m)$, and thus the one-form symmetry is broken to the diagonal one-form symmetry $B^e=B^m$. This is expected since the monopole acquires electric charge \cite{Witten:1979ey,Aharony:2013hda,Barkeshli:2022edm}.

The unbroken one-form symmetry has a mixed anomaly with the broken one-form symmetry due to the coupling $\pi\int b\cup B^m$ and the condition $db=-dB^e/2$ for $w_2^{PSO}$ to be $\mathbb{Z}_4$ cocycle. Thus the unbroken one-form symmetry fractionalizes. The unbroken one-form symmetry has nontrivial SPT.
    
\item If $n$ is even, we substitute $w_2^{(1)}=b$ that
corresponds to $w_2^{SO}$, and $w_2^{(2)}=B^e$, with the relation $db=-dB^e/2$ mod 2.
The low energy theory is
\begin{equation}
    n\pi\int b\cup \left(B^e+w_2\right)+2\pi \frac{nk}{8}\int {\cal P}(B^e)+\pi\int b\cup B^m=2\pi \frac{nk}{8}\int {\cal P}(B^e)+\pi\int b\cup B^m~,
\end{equation}
where the first term on the left hand side is trivial for even $n$. Thus the low energy theory is the topological order of $\mathbb{Z}_2$ gauge theory with boson particle, and the magnetic one-form symmetry is broken. The electric one-form symmetry is unbroken and not fractionalizes. The unbroken one-form symmetry has nontrivial SPT $2\pi \frac{nk}{8}\int {\cal P}(B^e)$.
    
\end{itemize}

Thus for odd $n$, the broken one-form symmetry in the two phases belongs to different subgroup of the electric and magnetic one-form symmetry. This is analogous to DQCP.

\subsubsection{Axial 0-form symmetry at massless point}

 When the fermions are massless, classically there is $U(1)_A$ axial symmetry, which is broken to a subgroup by the ABJ anomaly. When the axial symmetry is nontrivial such that the fermion mass term flips sign under the symmetry, the axial symmetry implements transformation between symmetry breaking unbroken phases. If the two phases have different topological orders, the axial symmetry is a non-invertible 0-form symmetry.

\paragraph{Example: $SO(3)$ gauge theory}

Consider $SU(2)$ gauge theory with two adjoint fermions.
At the massless point there is $\mathbb{Z}_4$ axial symmetry.
The axial symmetry has mixed anomaly with the electric one-form symmetry
\begin{equation}
    \frac{2\pi}{4}\int A\cup {\cal P}(B)~,
\end{equation}
where $A$ is the gauge field for the axial symmetry, and $B$ is the gauge field for the electric one-form symmetry.

If we gauge the one-form symmetry by promoting $B$ to be dynamical $b$ to obtain $SO(3)$ gauge theory with two vector fermions, the axial symmetry at the massless point becomes non-invertible \cite{Kaidi:2021xfk}.

\subsection{Chern-Simons matter theories in 2+1d}

Consider 2+1d $G$ gauge theory with continuous connected and simply connected $G$, and finite center $Z(G)\neq 1$.  

\paragraph{Gauge theories with bosons}

In gauge theory with adjoint Higgs fields, the theory has center electric one-form symmetry $Z(G)$. 

When the Higgs fields do not condense, the Chern-Simons level is zero, and the Wilson lines confine. Thus the electric one-form symmetry is unbroken.

When the Higgs fields condense, the gauge group can be broken to $Z(G)$, where the Wilson lines are deconfined and the electric one-form symmetry is spontaneously broken.

\paragraph{Gauge theories with fermions}

We couple the gauge field to fermion in the adjoint representation.  Under fermion mass transition, the Chern-Simons term of the gauge field can change from 0 to the dual Coxeter number $k=h^\vee$. In Chern-Simons theory $G_k$, the anyons are deconfined and the one-form symmetry is spontaneously broken. On the other hand, when the Chern-Simons level is zero, the Wilson lines confine and the one-form symmetry is unbroken. This is a standard Landau type symmetry breaking transition.

\section{Generalized SPT Transitions }
\label{sec:SPT}

In this section we will discuss models that exhibit SPT transitions of generalized symmetries, i.e. across the transition the theory to different SPT phases with generalized symmetries. The phase diagram is sketched in Fig.~\ref{sec:SPT}.

\begin{figure}[t]
    \centering
    \includegraphics[width=0.4\linewidth]{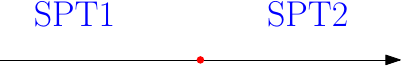}
    \caption{Phase diagram for SPT transition. The symmetries are unbroken, the two phases are trivially gapped.}
    \label{fig:SPT}
\end{figure}

\subsection{Gauge theories with bosons}

Consider gauge theory with gauge group $G$ and theta angle $\theta=2\pi k$ for integer $k$ such that the monopoles are confined. 
When the Higgs fields do not condense, the electric and magnetic one-form symmetries are unbroken, and the theory flows to SPT phases depending on $k$.

When the Higgs fields condense and break the gauge group to $K$, the SPT phase depends on $K$. If $K$ is the trivial group, then this gives trivial SPT. If $K$ is connected continuous group, the electric one-form symmetry is unbroken such that the theta angle inherited from the $G$ gauge theory also renders the monopoles to confine, 
the theory flows to SPT for the electric and magnetic one-form symmetry.

By comparing the SPTs, the Higgs transitions can realize SPT transitions for unbroken electric and magnetic one-form symmetries.

\subsubsection{Example: $SU(N_1N_2)/\mathbb{Z}_{N_2}$ gauge theory}

Consider $SU(N_1N_2)/\mathbb{Z}_{N_2}$ gauge theory with Higgs fields that break the gauge group completely. We turn on theta angle $\theta=2\pi k$ with $k$ to be determined.
The theory has $\mathbb{Z}_{N_2}$ magnetic one-form symmetry.

When the Higgs fields condense, the theory flows to trivial SPT. 
When the Higgs fields do not condense, the theory flows to the low energy two-form gauge theory 
\begin{equation}
    2\pi\frac{kN_1(N_1N_2-1) }{2N_2}\int {\cal P}(b)+\frac{2\pi}{N_2}\int b\cup B~,
\end{equation}
where $b,B$ are dynamical and background $\mathbb{Z}_{N_2}$ two-form gauge fields.
For the theory to be invertible, $\gcd(kN_1,N_2)=1$.
Then we can integrate out $b$ to give the SPT phase
\begin{equation}
    \Delta S_\text{SPT}=\frac{(kN_1)^{-1}}{2N_2}\int {\cal P}(B)~,
\end{equation}
where $(kN_1)^{-1}$ is a representative of the inverse of $kN_1$ in mod $N_2$.
This is the jump of SPT across the Higgs transition. As long as $\gcd(kN_1,N_2)=1$ this is a nontrivial SPT transition.

We note that for $N_1=1,N_2=2$, $k=1$, the model is $SO(3)$ gauge theory with $\theta=2\pi$, and the SPT transition is also discussed in \cite{Hsin:2021qiy}.

\subsubsection{Example: $PSO(4n)$ gauge theory}

\paragraph{Breaks gauge group completely.}

Consider $PSO(4n)$ gauge theory with Higgs fields that break the gauge group completely. We turn on theta angle $\theta=2\pi k$ with $k$ to be determined.
The theory has $\mathbb{Z}_{2}\times\mathbb{Z}_2$ magnetic one-form symmetry.

When the Higgs fields condense, the theory flows to trivial SPT. 
When the Higgs fields do not condense, for $nk=1,3$ mod 4 the theory flows to invertible two-form gauge theory:
 \begin{equation}
        \pm \frac{2\pi}{4}\int {\cal P}(b'^{(2)})\pm \frac{2\pi}{4}\int {\cal P}(b^{(1)})+\pi\int b^{(1)}\cup (B^{(1)}+B^{(2)})+\pi\int b'^{(2)}\cup B^{(2)}~,
    \end{equation}
    where $b^{(i)}$ are dynamical two-form $\mathbb{Z}_2$ gauge fields, $B^{(i)}$ are background two-form $\mathbb{Z}_2$ gauge fields for the magnetic one-form symmetry. We use the change of variables $b'^{(2)}=b^{(2)}+b^{(1)}$. In the above expression, the sign $\pm$ corresponds to $nk=1,3$ mod 4. 
    After integrating out $b^{(i)}$, we find the SPT phase 
    \begin{align}
\Delta S_\text{SPT}&=        \mp \frac{2\pi}{4}\int {\cal P}(B^{(1)}+B^{(2)})\mp\frac{2\pi}{4}\int {\cal P}(B^{(2)})\cr
&=\mp \frac{2\pi}{4}\int {\cal P}(B^{(1)})+\pi \int B^{(1)}\cup B^{(2)}+\pi\int B^{(2)}\cup B^{(2)}~.
    \end{align}
Thus the Higgs transition gives nontrivial SPT transition.

\paragraph{Breaks gauge group to $SU(2n)/\mathbb{Z}_2$.}

Consider $PSO(4n)$ gauge theory with Higgs fields that break the gauge group to $SU(2n)/\mathbb{Z}_2$ via the embedding $SU(2n)\subset SO(4n)$.
Again we will focus on the magnetic one-form symmetry.
The $SU(2n)/\mathbb{Z}_2$ gauge theory has $\mathbb{Z}_2$ magnetic one-form symmetry that corresponds to $B^{(2)}$.
In the Higgs phase, the low energy theory is
\begin{equation}
    2\pi\frac{k(2n-1)}{4n} n^2\int {\cal P}(b)+\pi\int b\cup B^{(2)}~.
\end{equation}
For $nk=1,3$ mod 4, this gives
\begin{equation}
    \pm\frac{2\pi (2n-1)}{4}\int {\cal P}(b)+\pi\int b\cup B^{(2)}~.
\end{equation}
After integrating out $b$, we obtain the SPT
\begin{equation}
    \mp(2n-1)\frac{2\pi}{4}\int {\cal P}(B^{(2)})~.
\end{equation}
The difference of SPT across the Higgs transition is
\begin{equation}
    \Delta S_\text{SPT}=\mp \frac{2\pi}{4}\int {\cal P}(B^{(1)})
    +\pi\int B^{(1)}\cup B^{(2)}\mp (2n-1)\frac{2\pi}{4}\int {\cal P}(B^{(2)})~.
\end{equation}
Thus as long as $nk=1,3$ mod 4, the Higgs transition is a nontrivial SPT transition.

\subsubsection{Example: $PSO(4n+2)$ gauge theory}

Consider $PSO(4n+2)$ gauge theory with Higgs fields that break the gauge group completely. We turn on theta angle $\theta=2\pi k$ with $k$ to be determined.  The theory has $\mathbb{Z}_4$ magnetic one-form symmetry.

When the Higgs fields condense, the theory is trivial SPT. When the Higgs fields do not condense, the low energy theory is
    \begin{equation}
        \frac{2\pi (2n+1)k}{8}\int {\cal P}(b)+\frac{2\pi}{4}\int b\cup B~,
    \end{equation}
where  $b$ is a dynamical two-form $\mathbb{Z}_4$ gauge field, while $B$ is the background two-form $\mathbb{Z}_4$ gauge field for the magnetic symmetry. 
When $k=1,3$ mod 4, the theory is invertible. After integrating out $b$, this gives the SPT phase
    \begin{equation}
\Delta S_\text{SPT}=        -\frac{2\pi (2n+1)k}{8}\int {\cal P}(B)~,
    \end{equation}
Thus as long as $k=1,3$ mod 4, the Higgs transition is a nontrivial SPT transition.

\subsubsection{Example: $PSp(N)$ gauge theory}

Consider $PSp(N)$ gauge theory with Higgs fields that break the gauge group completely. We turn on theta angle $\theta=2\pi k$ with $k$ to be determined.  The theory has $\mathbb{Z}_2$ magnetic one-form symmetry.

When the Higgs fields condense, the theory is trivial SPT. When the Higgs fields do not condense, the low energy theory is invertible for $Nk=1,3$ mod 4: after integrating out the dynamical two-form gauge field, this gives the SPT
    \begin{equation}
\Delta S_\text{SPT}=        \mp \frac{2\pi}{4}\int {\cal P}(B)~,
    \end{equation}
    where $\mp$ corresponds to $Nk=1,3$ mod 4, and $B$ is the background two-form $\mathbb{Z}_2$ gauge field for the magnetic symmetry. Thus the as long as $Nk=1,3$ mod 4, the Higgs transition is a nontrivial SPT transition.

\subsection{Gauge theories with fermions}

When the fermion mass change sign, the gauge field coupled to the fermion can acquire different topological terms. Such topological terms can lead to change of SPT phases. 

The change of theta angle $\theta=2\pi k_1$ to $\theta=2\pi k_2$  with integers $k_1,k_2$ due the fermion mass transition can change the SPT phase of electric one-form symmetry \cite{Gaiotto:2014kfa}. In addition, the theta angles with different $k_1,k_2$ can give different SPT phases for the magnetic one-form symmetry.

We remark that the SPT phases with one-form symmetries in 3+1d do not require additional time-reversal symmetry, unlike the theta terms for 0-form symmetries that can be smoothly tuned to zero without time-reversal symmetry. Thus we do not need to impose time-reversal symmetry in the discussion here. For simplicity, we focus on the positive and negative fermion masses; there can be other transitions when the asymptotic phases correspond to different complex masses.

\subsubsection{Example: $SU(N)$ gauge theory}

Let us review the SPT transition in $SU(N)$ gauge theory with adjoint fermions \cite{Gaiotto:2014kfa}. The theory has $\mathbb{Z}_N$ electric center one-form symmetry.

The fermion transition can change the theta angle by $\Delta \theta=\pi NN_f$ for $N_f$ flavors of fermions, which leads to SPT transition with the jump
\begin{equation}
    \Delta S_\text{SPT}=2\pi\frac{(N-1)}{2N}\frac{NN_f}{2}\int {\cal P}(B)=2\pi \frac{N_f(N-1)}{4}\int {\cal P}(B)~.
\end{equation}
For $N_f(N-1)\neq 0$ mod 4 and even $N$ this is a nontrivial SPT transition.

\subsubsection{Example: $SO(2n)_-$ gauge theory}

Consider $SO(2n)_-$ gauge theory with discrete theta angle topological term \cite{Aharony:2013hda,Gaiotto:2014kfa,Hsin:2018vcg,Ang:2019txy}
\begin{equation}
    \frac{2\pi}{4}\int {\cal P}(w_2^{SO})+\pi\int w_2^{SO}\cup B~,
\end{equation}
where $B$ is the background for the $\mathbb{Z}_2$ magnetic one-form symmetry. We will consider fermions where the fermion parity is identified with the center of the gauge group. Thus the theory is effectively bosonic. Moreover, the theory has time-reversal symmetry \cite{Hsin:2021qiy}.

When the fermion has positive mass, the theory is trivially gapped, and it is equivalent to the SPT
\begin{equation}
    -\frac{2\pi}{4}\int {\cal P}(B)~.
\end{equation}

When the fermion has negative mass, the theory has additional 
theta angle $\theta=2\pi k$ for integer $k$, which changes the discrete theta angle into
\begin{equation}
    \frac{2\pi (2k+1)}{4}\int {\cal P}(w_2^{SO})+\pi\int w_2^{SO}\cup B~,
\end{equation}
which is also trivially gapped, and it is equivalent to the SPT
\begin{equation}
    -\frac{2\pi(2k+1)}{4}\int {\cal P}(B)~.
\end{equation}
Thus the jump of SPT across the fermion mass transition is
\begin{equation}
    \Delta S_\text{SPT}=\pi k\int B\cup B=\pi k\int B\cup w_2~,
\end{equation}
which is nontrivial for odd $k$. Thus as long as $k$ is odd, e.g. two Dirac fermions in the vector representation, the fermion mass transition is a nontrivial SPT transition.

The discussion for other gauge groups is similar, and we will not repeat it here.

\subsubsection{Axial 0-form symmetry at massless point}

When the fermions are massless, classically the action has $U(1)_A$ axial 0-form symmetry, and it is broken by non-Abelian ABJ anomaly to a subgroup determined by the Dynkin index such that the corresponding change of theta angle is trivial. The fact that across the massless point there is jump of SPT for the one-form symmetries indicates a nontrivial mixed anomaly between the remanent axial symmetry and the one-form symmetry at the massless point, where the axial transformation ``eats'' the difference of SPT phase \cite{Hsin:2019fhf,Dumitrescu:2023hbe}.

\paragraph{Example: $SU(N)$ gauge theory}

Consider $SU(N)$ gauge with adjoint Weyl fermion for even $N$. The theory has $\mathbb{Z}_N$ electric one-form symmetry.  At massless point there is $\mathbb{Z}_{2N}$ axial symmetry.
The fermion mass transition induces the jump of theta angle $\Delta \theta=\pi N$, which is equivalent to jump of SPT
\begin{equation}
    \Delta S_\text{SPT}=\frac{2\pi (N/2)(N-1)}{2N}\int {\cal P}(B)=\frac{2\pi(N-1)}{4}\int {\cal P}(B)~.
\end{equation}

The axial symmetry has mixed anomaly with the one-form symmetry \cite{Gaiotto:2014kfa}: the axial transformation shifts the theta angle by $2\pi$, which is a nontrivial SPT. The anomaly is described by the 4+1d term 
\begin{equation}
    \frac{2\pi(N-1)}{2N}\int A\cup {\cal P}(B)~,
\end{equation}
where $A$ is the background for the axial symmetry.

 \paragraph{Example: $Spin(4n)$ gauge theory}

Consider $Spin(4n)$ gauge theory with vector Dirac fermion. The theory has $\mathbb{Z}_2$ center electric one-form symmetry.
The massless point has axial symmetry $\mathbb{Z}_4$, where the Dirac mass term flips sign and gives a jump of theta angle by $2\pi$.
The fermion parity is identified with a center of $Spin(4n)$, and the theory is effectively bosonic. 

In the $Spin(4n)$ gauge theory, the fermion mass transition changes the SPT by
\begin{equation}
    \pi\int B\cup B+\pi\int B\cup w_2+\frac{2\pi (n/2)}{4}\int {\cal P}(w_2)=\frac{2\pi n}{4}\int {\cal P}(w_2)~,
\end{equation}
where $B$ is the background for the electric one-form symmetry.
Thus the axial symmetry has mixed anomaly 
\begin{equation}
    \frac{2\pi n}{4}\int A\cup {\cal P}(w_2)~.
\end{equation}

We remark that such symmetry is not always present at the SPT transitions, such as the transitions described by gauge theory with bosons.

\section{Generalized SET Transitions}
\label{sec:SET}

\begin{figure}[t]
    \centering
    \includegraphics[width=0.4\linewidth]{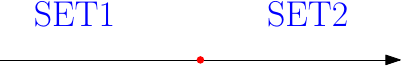}
    \caption{Phase diagram for SET transitions. The symmetries are unbroken. We will focus on the case that the two phases have the same topological orders if symmetries are ignored (including the cases of Lorentz symmetry fractionalization \cite{Hsin:2019gvb}).}
    \label{fig:SET}
\end{figure}

In this section we will discuss transitions separated topological phases with different symmetry fractionalizations. We will focus on the cases of the same topological orders but different symmetry fractionalizations. The phase diagram is sketched in Fig.~\ref{fig:SET}. 

\subsection{Gauge theory with bosons}

When fractionalizations correspond to anomalies, it cannot occur in the same model due to anomaly matching. Thus we should focus on fractionalizations that do not give anomalies for exact symmetries.
Consider $G$ gauge theory for continuous connected $G$ with Higgs fields such that there is no electric center one-form symmetry but only magnetic one-form symmetry $\pi_1(G)$. We will set up the models such that the gauge group can be broken to two different isomorphic finite subgroups $K_1\cong K_2:=K$ under different Higgs fields. The two Higgs phases have the same fintie group $K$ gauge theory topological order, but the magnetic one-form symmetries can have different fractionalizations in these Higgs phases.
Then, we can construct a path in the phase diagram between these Higgs phases, and the path realizes transitions between different SET for $K$ gauge theory.

\subsubsection{Example: $SO(N)$ gauge theory with two Higgs phases of different SETs}

\begin{figure}[t]
    \centering
    \includegraphics[width=0.7\linewidth]{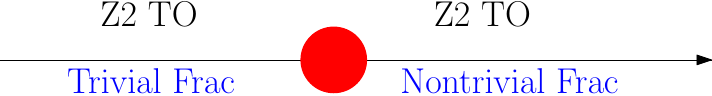}
    \caption{A 1d path in the $SO(3)$ gauge theory phase diagrams between two Higgs phases with $\mathbb{Z}_2$ gauge groups in 3+1d. The path describes SET transition that separates $\mathbb{Z}_2$ gauge theories with different fractionalizations of the magnetic $\mathbb{Z}_2$ one-form symmetry of the $SO(3)$ gauge theory. The $\mathbb{Z}_2$ gauge theories have emergent one-form symmetry and two-form symmetry that are absent in the UV $SO(3)$ gauge theory model.}
    \label{fig:SO3Z2SET}
\end{figure}

Consider $SO(N)$ gauge theory with magnetic one-form symmetry. We will discuss two Higgs phases where the magnetic one-form symmetry is unbroken, both phases have the $\mathbb{Z}_2$ gauge theory topological order, but they differ by how the magnetic one-form symmetry fractionalizes. 

We will illustrate the construction using $SO(3)$.
We can break $SO(3)$ in two routes:
\begin{itemize}
    \item Break $SO(3)$ to Cartan $SO(2)\subset \mathbb{Z}_2$ and then to $\mathbb{Z}_2$. In this route, the $\mathbb{Z}_2$ magnetic one-form symmetry fractionalizes.
    
To see this, we note that the $\mathbb{Z}_2$ magnetic flux is related to the Cartan gauge field $a$ of $SO(3)$ as
\begin{equation}
    w_2^{SO}=\frac{da}{2\pi} \text{ mod }2~.
\end{equation}
Thus if we further break the Cartan $U(1)$ gauge group to $\mathbb{Z}_2$, the resulting $\mathbb{Z}_2$ gauge theory has the coupling
\begin{equation}
    \frac{da}{2\pi}B+\frac{2}{2\pi}dab~,
\end{equation}
where $b$ is dual to the phase of the Higgs field, and $B$ is the background field for the magnetic one-form symmetry. This is exactly the setup in section \ref{sec:U1example}, and the magnetic one-form symmetry fractionalizes on the $\mathbb{Z}_2$ vortex loops.

    \item Break $SO(3)$ to $O(2)$ and then to $\mathbb{Z}_2^{\cal C}$ given by the orientation reversal of $O(2)$. In this route, the magnetic one-form symmetry does not fractionalize.
This is because the magnetic flux of $O(2)$ gauge field, which couples to $B$, does not depend on the $\mathbb{Z}_2^{\cal C}$ gauge field.
    
\end{itemize}
In the $\mathbb{Z}_2$ topological orders of the two Higgs phases, the $\mathbb{Z}_2$ center one-form symmetry is emergent and it is absent in the UV $SO(3)$ gauge theory which has trivial center.
We sketch the 1d path in the phase diagram in Fig.~\ref{fig:SO3Z2SET}.

For more general $SO(N)$, we can break the gauge group to a Cartan $SO(2)$ and then to $\mathbb{Z}_2$ in the first route, and break the gauge group to $O(N-1)$ and to $\mathbb{Z}_2^{\cal C}$ in the second route. The conclusion is the same: they are different SET phases with the same $\mathbb{Z}_2$ gauge theory topological order.
The discussion can be generalized to other continuous gauge groups.

\subsubsection{Example: $SU(2n)$ gauge theory Higgsed to $SO(2n')$}

Consider $SU(2n)$ gauge theory with Higgs fields that can break the gauge group to $SO(2n')$ with $n'\leq n$. The theory has $\mathbb{Z}_2$ electric center one-form symmetry. 

When the Higgs fields do not condense, the theory is trivially gapped and the electric one-form symmetry is unbroken.

When the Higgs fields condense, the gauge group is broken to $SO(2n')$, the theory flows to $\mathbb{Z}_2$ gauge theory topological order with broken emergent $\mathbb{Z}_2$ magnetic one-form symmetry, and unbroken $\mathbb{Z}_2$ one-form symmetry.

If $n'$ is odd, the unbroken electric one-form symmetry
fractionalizes on the $\mathbb{Z}_2$ loop excitation due to the mixed anomaly with the emergent magnetic one-form symmetry. Thus the Higgs transition is a phase transition between a trivial SPT and nontrivial SET for the $\mathbb{Z}_2$ center one-form symmetry.

Similarly, if the gauge group is Higgsed to $SO(2n')$ with even $n'$, there is another emergent magnetic one-form symmetry that is broken by $\mathbb{Z}_2$ topological order, and the unbroken electric one-form symmetry does not fractionalize on the loop excitations.

\subsubsection{Example: $SO(2n)$ gauge theory with $\theta=2\pi$}

Consider $SO(2n)$ gauge theory with $\theta=2\pi$ for $n\geq 2$ and adjoint Higgs fields that can break the gauge group to $\mathbb{Z}_2$.
The theory has $\mathbb{Z}_2$ magnetic one-form symmetry and $\mathbb{Z}_2$ electric one-form symmetry.

When the Higgs fields do not condense, the theory flows to $\mathbb{Z}_2$ gauge theory with fermion charge from deconfined fermion monopoles. The loop excitation carries charge $1/2$ under the electric one-form symmetry.

When the Higgs fields condense and the gauge group is broken to center $\mathbb{Z}_2$, the low energy theory flows to $\mathbb{Z}_2$ gauge theory with boson charges from the deconfined Wilson line. 
The electric one-form symmetry is broken, and the magnetic one-form symmetry is unbroken and fractionalizes on the vortex loop excitations.

There is no one-form symmetry remains unbroken across the transition, and thus we cannot discuss the SET transition for one-form symmetry. On the other hand, the topological order changes by Lorentz symmetry fractionalization across the transition where the boson charge becomes fermion charge \cite{Hsin:2019gvb}, and thus the transition is an SET transition for Lorentz symmetry.

\subsection{Gauge theory with fermions}

The fermion mass transition changes the subgroup of magnetic symmetry that is unbroken, as well as SPT phases for one-form symmetries. On the other hand, it does not change the fractionalization of one-form symmetries.\footnote{Note that while gauging  0-form symmetry in a 0-form SPT can change the fractionalization of 0-form symmetry, gauging one-form symmetries in the 3+1d one-form SPT phases does not change the fractionalization.}

However, shift of theta angles induced by fermion mass transition can change the Lorentz symmetry fractionalization \cite{Hsin:2019gvb}, where the statistics of deconfined excitations \cite{Kobayashi:2024dqj} change across the transition. This happens when the jump of theta angle can be expressed in terms of discrete theta angle that is $\mathbb{Z}_2$ valued on general orientable non-spin manifolds:
\begin{equation}
    \pi\int w_2^G\cup w_2^G=\pi\int w_2^G\cup w_2~,
\end{equation}
where $w_2$ is the second Stiefel-Whitney class of the tangent bundle \cite{milnor1974characteristic}. The coupling indicates that the spins of the monopoles that transforms under the $\mathbb{Z}_2$ subgroup magnetic symmetry $(-1)^{\int w_2^G}$ are shifted by $1/2$ \cite{Hsin:2019gvb}. If the monopoles are deconfined, then the fermion mass transition is an SET transition for closely related topological orders with different Lorentz symmetry fractionalizations.

\subsubsection{Example: $SO(N)$ gauge theory}

Consider $SO(N)$ gauge theory with vector fermions and zero theta angle. The fermion mass transition can induce the shift of theta angle
\begin{equation}
    \Delta\theta=2\pi k~.
\end{equation}
We will focus on the case of odd $k$. Then the shift of theta angle is equivalent to the discrete theta angle
\begin{equation}
    \pi\int w_2^{SO}\cup w_2^{SO}=\pi\int w_2^{SO}\cup w_2~,
\end{equation}
where $w_2$ is the second Stiefel Whitney class of the tangent bundle. 

When the fermion mass is positive, the theory flows to $\mathbb{Z}_2$ gauge theory with boson charge. When the fermion mass is negative, the theory flows to $\mathbb{Z}_2$ gauge theory with fermion charge due to the coupling to $w_2$ \cite{Hsin:2019gvb}. Such change of fractionalization gives SET transition for Lorentz symmetry. Throughout the transition, the electric one-form symmetry (present for even $N$) is unbroken, and the $\mathbb{Z}_2$ magnetic one-form symmetry is broken.

\section{Outlook}
\label{sec:outlook}

There are several future directions.
First, 
while most of the transitions discussed here are likely of first order such as the Higgs transitions, it would be interesting to find continuous transitions using the method. This requires imposing additional 0-form symmetries (invertible or non-invertible \cite{Shao:2023gho,Schafer-Nameki:2023jdn}), such as axial symmetry for massless fermions or flavor symmetries, to reduce relevant operators which are automatically singlet under higher-form symmetries. Otherwise, one needs to fine tune to obtain critical points.

Second, it would be interesting to exam the properties of defects and excitations at the transitions, such as domain walls from varying the parameters and lower dimensional defects from shrinking the domain walls. For example, the domain walls from Higgs transitions can be analyzed using similar method as in e.g. \cite{Cordova:2025eim}. In the SPT transitions, the domain walls have anyons that match the anomalous one-form symmetry corresponds to the jump of SPTs \cite{Hsin:2018vcg}. Similarly, in the SET transitions, the domain walls are nontrivial to account for the fractional charges carried by loop excitations \cite{PhysRevX.6.011034,Hsin:2019fhf}. The details will be discussed in a separate work.

Finally, the discussions can be generalized to non-linear sigma models instead of gauge theories. Non-linear sigma models have (invertible or non-invertible) higher-form symmetries associated with topological terms of the sigma model fields or sigma model fields coupled to TQFTs on submanifolds \cite{Hsin:2022heo,Chen:2022cyw,Hsin:2022iug,Pace:2023kyi,Pace:2023mdo,DelZotto:2024arv}. 
Under perturbations with sigma model fields, these symmetries are preserved, but the magnetic defects can become confined when the new vacua pinned by the perturbations do not support the boundary conditions for the magnetic defects. It would interesting to exam the phase transitions of sigma models with generalized symmetries.

\appendix

\section{Continuous and Discrete Theta Angles}
\label{sec:thetaangle}

In this appendix we summarize the relation between theta angles $\theta=2\pi k$ for integer $k$ and discrete theta angles in 3+1d for different continuous gauge groups. For more detail, see e.g. \cite{Gaiotto:2014kfa,Hsin:2018vcg,Cordova:2019uob}. The theta angles correspond to the following discrete theta angles:
\begin{itemize}
    \item When $G=SU(N)/\mathbb{Z}_\ell$ for $\ell|N$, it is
    \begin{equation}\label{eqn:discthetaSU}
        2\pi \frac{k(N-1)}{2N}(N/\ell)^2\int {\cal P}(w_2^{SU(N)/\mathbb{Z}_\ell})=2\pi \frac{p}{2\ell}\int {\cal P}(w_2^{SU(N)/\mathbb{Z}_\ell}),\quad p=kN(N-1)/\ell~,
    \end{equation}
    where $w_2^{SU(N)/\mathbb{Z}_\ell}$ is the $\mathbb{Z}_\ell$ value two-form for the discrete magnetic flux.

    \item When $G=SO(N)=Spin(N)/\mathbb{Z}_2$, it is
\begin{equation}\label{eqn:discthetaSO}
    k \pi\int w_2^{SO}\cup w_2^{SO}~,
\end{equation}
where $w_2^{SO}$ is the $\mathbb{Z}_2$ two-form for the discrete magnetic flux.

    \item When $G=PSO(4n+2)=Spin(4n+2)/\mathbb{Z}_4$, it is
    \begin{equation}\label{eqn:discthetaSO4n+2}
        2\pi \frac{k(4n+2)}{16}\int {\cal P}(w_2^{PSO})=2\pi\frac{k(2n+1)}{8}\int  {\cal P}(w_2^{PSO})~,
    \end{equation}
    where $w_2^{PSO}$ is the $\mathbb{Z}_4$ value two-form for the discrete magnetic flux.

    \item When $G=PSO(4n)=Spin(4n)/(\mathbb{Z}_2\times\mathbb{Z}_2)$, it is
    \begin{equation}\label{eqn:discthetaSO4n}
       \pi k \int w_2^{(1)}\cup w_2^{(1)}+\pi k\int w_2^{(1)}\cup w_2^{(2)}+
       2\pi \frac{k(4n)}{16}\int {\cal P}(w_2^{(2)})~,
    \end{equation}
    where $w_2^{(1)},w_2^{(2)}$ are $\mathbb{Z}_2\times\mathbb{Z}_2$ value two-form for the discrete magnetic fluxes associated with the quotient $SO(4n)=Spin(4n)/(\mathbb{Z}_2\times\mathbb{Z}_2)$.

    \item When $G=PSp(N)=Sp(N)/\mathbb{Z}_2$, it is
    \begin{equation}\label{eqn:discthetaSp}
        2\pi \frac{Nk}{4}\int {\cal P}(w_2^{PSp})~,
    \end{equation}
where $w_2^{PSp}$ is the $\mathbb{Z}_2$ two-form gauge field for the discrete magnetic flux.
Here the notation of $Sp(N)$ is such that $Sp(1)=SU(2),Sp(2)=Spin(5)$. 
    
\end{itemize}

\section*{Acknowledgment}

P.-S.H. is supported by Department of Mathematics King's College London. P.-S. H. thanks Kavli Institute for Theoretical Physics for hosting the program ``Generalized Symmetries in Quantum Field Theory: High Energy Physics, Condensed Matter, and Quantum Gravity,'' during which part of the research was completed.
This research was supported in part by grant NSF PHY-2309135 to the Kavli Institute for Theoretical Physics (KITP).

\bibliographystyle{utphys}
\bibliography{biblio}

\end{document}